\documentstyle[preprint,aps,prb]{revtex}

\newcommand{\mb}[1]{ {\mbox{\boldmath{$#1$}}}  }

\begin{document}
\draft

\title{Van Hove Singularity and Superconductivity in Disordered
Sr$_2$RuO$_4$
}

\author{G. Litak}
\address{Department of Mechanics,
 Technical University of Lublin\\ 
Nadbystrzycka 36, PL--20-618  Lublin, Poland}

\author{ J.F. Annett and B.L. Gy\"{o}rffy}
\address{H.H. Wills Physics Laboratory, University of
Bristol \\ Tyndall Ave, Bristol BS8 1TL, United Kingdom
}

\date{\today}
\maketitle

\begin{abstract} 
On the basis of a simple model we analyse the influence of disorder on
critical temperature T$_C$ in p--wave
superconductors. The disorder is treated by means of  the Coherent
Potential Approximation (CPA) and we focus our attention on 
the effect of a van Hove
singularity 
near Fermi energy E$_F$. 
For the appropriate values of its parameters our model
reproduces the experimentally found behaviour of 
Sr$_2$RuO$_4$.
\end{abstract}

\section{Introduction}

The perovskite structure of Strontium Ruthenate, Sr$_2$RuO$_4$
is very similar to
that of HTS
copper oxides. However, 
its superconducting transition temperature T$_C$ is relatively low 
(T$_C$ $\approx$ 1 K) [1]. Nevertheless, recent reports indicate that its
Cooper pairs are not of the usual s--wave symmetry.
In fact they suggest that this material features  
triplet pairing and is a
superconducting analogue of the  $^3$He superfluid system [1-4].    
Clearly, the possibility of exotic  pairing
engenders interest in  the effects of disorder on the
superconducting properties.
Moreover, studies of the electronic structure 
[2,3] have identified an extended van Hove singularity  
close to the Fermi energy E$_F$, and therefore
one may wonder whether the van
Hove  scenario could lead to a rise in T$_C$ with doping.
Evidently, since doping  the system always increases the disorder
one should investigate both aspects simultaneously.

\section{The Model}

We base our discussion on  the extended negative U Hubbard Hamiltonian: 

%eq1
\begin{equation}
\label{eq:H}
H=\sum_{ij \sigma} t_{ij}c^{\dagger}_{i \sigma} c_{j \sigma} +
\frac{1}{2} \sum_{ij} U_{ij} \hat n_i \hat n_j -
\sum_{i}(\mu-\varepsilon_i) \hat n_i,
\end{equation}
where $\hat n_i=\hat n_{i \uparrow}+\hat n_{i \downarrow}$ and
$\hat n_{i \sigma}$ is the usual, site occupational number
operator $c^{\dagger}_{i \sigma} c_{i \sigma}.$
Evidently the above $\hat n_i$ is the charge operator on  site labelled
$i$, $\mu$ is the
chemical potential, which at $T=0$ is equal to Fermi energy $E_F$.
Disorder is introduced into the problem by allowing the local
site
energy $\varepsilon_i$ to vary randomly from site to site.
Finally, $c^{\dagger}_{i
\sigma}$ and $c_{i \sigma}$ are the Fermion
creation and annihilation operators for an electron
 on site $i$ with spin $\sigma$,
$t_{ij}$ is the amplitude for
hopping from site $j$ to site $i$ and $U_{ij}$ is the attractive interaction ($i\ne j$)
which causes superconductivity.

In the Hartree-Fock-Gorkov approximation the equation for the Green's
function
$\mb G(i,j;
\imath \omega_n)$, corresponding to the Hamiltonian in Eq. 1, is given by:

%eq2
\begin{equation}
\sum_l \left[  \begin{array}{cc} (\imath \omega_n + \mu - \varepsilon_i) 
\delta_{il} + t_{il} & \Delta_{il}
\\ \Delta_{il}^*  & (\imath \omega_n - \mu + \varepsilon)
\delta_{il} - t_{il}  
\end{array}
\right] \mb G(l,j; \imath \omega_n) = \mb 1 \delta_{ij},
\end{equation}
 where $\omega_n$ is Matsubara frequency.Let us define the  random
potential $\mb
V^{\varepsilon_i}$ by: 

%eq3
\begin{equation}
\mb V^{\varepsilon_i} = \left[ \begin{array}{cc} \varepsilon_i & 0 \\ 0 & -\varepsilon_i
\end{array}
\right],
\end{equation}
where $\varepsilon_i$  is uniformly distributed on the energy interval
$[-\frac{\delta}{2},\frac{\delta}{2}]$. The Green's function for an 
impurity, described by
$\mb V^{\varepsilon_i}$ in Eq. 3, embedded in the  medium, described by
$\mb \Sigma (\imath \omega_n)$  is given by:
%eq4
\begin{equation}
\mb G^{\varepsilon_i} (i,i, \imath \omega_n) = \{ \mb 1- \mb G^C  (i,i, \imath \omega_n) [ \mb V^{\varepsilon_i} - \mb \Sigma(\imath \omega_n)]\}^{-1}
 \mb G^C  (i,i, \imath \omega_n),
\end{equation}   
Following the usual CPA procedure we demand that the coherent potential
Greens function $ \mb G^C(i,i; \imath \omega_n)= ( \imath \omega_n -
\epsilon_k - \mb \Sigma(\imath \omega_n))^{-1}$ satisfy the relation:

 %eq5
\begin{equation} \mb
G^C  (i,i, \imath \omega_n)= \langle \mb G^{\varepsilon_i}  (i,i, \imath \omega_n) \rangle =
\frac{1}{\delta} \int_{-\delta/2}^{\delta/2} {\rm d} \varepsilon_i~~  \mb G^{\varepsilon_i} (i,i, \imath \omega_n).
\end{equation}
Evidently, Eq. 5 completely determines, that is to say can be solved for,
$\mb \Sigma (\imath \omega_n)$.

Let us now proceed further with the CPA strategy [5] and determine the
averaged
Greens function matrix $\langle\mb{G}(i,j;\imath \omega_n)\rangle$ subject
to
the self consistency conditions:
%eq6
\begin{equation}
\label{eq:selfcon2}
\overline \Delta_{ij}= |U_{ij}| \frac{1}{\beta}
\sum_n {\rm e}^{\imath \omega_n \eta}
\langle G_{12} (i,j;\imath \omega_n) \rangle,~~~
\overline n= \frac{2}{\beta} \sum_n {\rm e}^{\imath \omega_n \eta}
\langle G_{11} (i,i;\imath \omega_n) \rangle.
\end{equation}

In this paper we assumed nearest neighbour electron hopping and pairing
on a two dimensional lattice.
In Figure 1$a$ and $b$ we have presented Fermi surfaces for $n=0.55$ and
$n=1$
respectively. The latter case correspond to the situation, where 
the Fermi Energy, $E_F$,
is located exactly at the van Hove singularity.

\section{Critical Temperature and Residual Resistivity}

The linearised gap equation for the critical temperature T$_C$ of p-wave
superconducting phase transition reads as follows [6]:

\begin{equation}
1 = |U| T_C \sum_n {\rm e}^{\imath \omega_n \eta} \frac{1}{N} \sum_{\vec
k} \frac{2 (\sin k_x)^2}{ (\imath
\omega_n - \epsilon_{\vec k} - \Sigma_{11}(\imath \omega_n)) (\imath
\omega_n + \epsilon_{\vec k} -
\Sigma_{22}(\imath \omega_n))}.
\end{equation}

A useful measure of disorder is the resistivity $\rho$. Thus we shall
study the relationship between $\rho$ and T$_C$.
 The Residual resistivity $\rho$ for low temperature can be obtained from
The Kubo--Greenwood formula. For the disordered
two dimensional systems at hand [7]:

%eq8
\begin{equation}
\rho= \left\{ 2 \frac{{\rm e}^2}{\pi \hbar c} \frac{1}{N} \sum_k 4 (\sin
k_x)^2
t^2 \left[ {\rm Im} 
G^C_{11}( \vec k, 0) \right]^2 \right\}^{-1},   
\end{equation}
where e  is the electron charge, $\hbar$ is Plank constant and $c$ is
the distance between RuO$_2$ planes. 

In short, we have solved the CPA equations (Eqs. 4,5) for various system
parameters (Eq. 1) and calculated both T$_C$ and residual 
resistivity $\rho$.

To illustrate how effective a van
Hove singularity can
be in raising T$_C$, in Fig. 2$a$ we present T$_C$, calculated for clean
systems and
normalised to its maximal value T$_C^{max}$,
 versus band filling $n$ for various values of $U/t$.
Clearly, T$_C$ is peaked at $n=1$, where the Fermi energy $E_F$ is
exactly at
the van Hove
singularity.
For small enough interaction $U$ it is enlarged   by a
factor of  7. Going further  we turn to our results for the disordered
case. Thus, in Fig. 2$b$, we plotted T$_C$ versus residual
resistivity $\rho$ as calculated by the CPA procedure described above.
The parameters $U/t=-0.702$ as well as
band filling $n=0.55$ were chosen so that the T$_C$ vs. $\rho$ curve 
reproduce the experiments [1].
Unlike the Born approximation limit, the CPA residual resestivity  is
dependent
on the strength of disordered
potential, $\delta$, nonlinearly. This is  illustrated in  Fig. 3$a$,
where
the different
curves correspond to different
band fillings $n$. The pronounced nonlinearity for n=1 is due to 
a van Hove singularity being
near $E_F$. As shown in Fig. 3b this give rise to an interesting upturn as
$\rho \rightarrow 0$ in the T$_C$ vs. $\rho$ plot.

\section{Remarks and Conclusions}

Our results confirm  that, similarly to d--wave superconductors [5], in
the case of p--wave paring
the
critical
temperature T$_C$ is very sensitive function of  nonmagnetic diagonal
disorder.
Nevertheless, they sugest that in Sr$_2$RuO$_4$ 
doping could lead to   higher value of critical temperature
T$_C$. 
Here we used uniform distribution of site energy levels $\varepsilon_i$ 
as the simplest model of disorder.
Clearly further study of the problem would include 
a more sophisticated impurity model,
and more realistic band structure.

\section*{Acknowledgements}

Authors would like to thank Dr A.M. Martin,  
Prof. K.I. Wysoki\'nski for helpful discussions and Prof. A.P. Mackenzie
for the experimental data.
This work has been partially supported by KBN grant No. 2P03B05015
and EPSRC grant  No. GR/L22454.

\begin{figure}[htb]
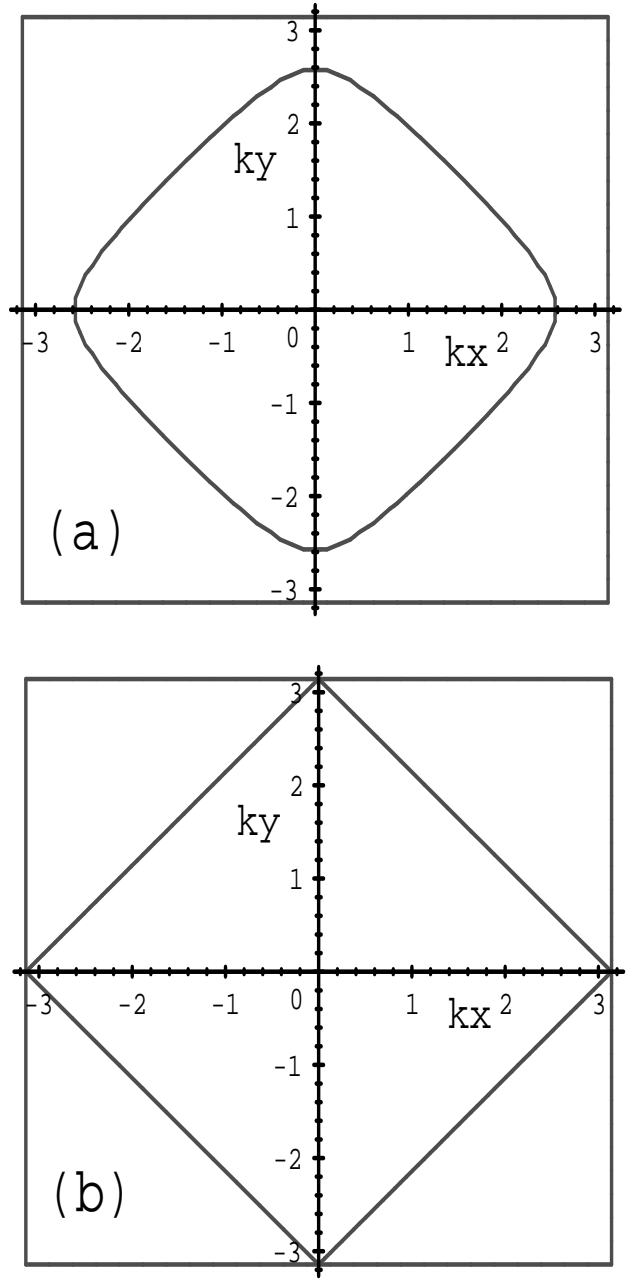

\caption[Fig. 1]{Fermi surfaces for the one band electron structure
with nearest neighbour hoping: $\epsilon_{\vec k} = -2 t (\cos k_x +
\cos k_y)$, and two different band fillings: $n=0.55$ ($a$),
 $n=1.00$ ($b$).}
\end{figure}

\begin{figure}[htb]
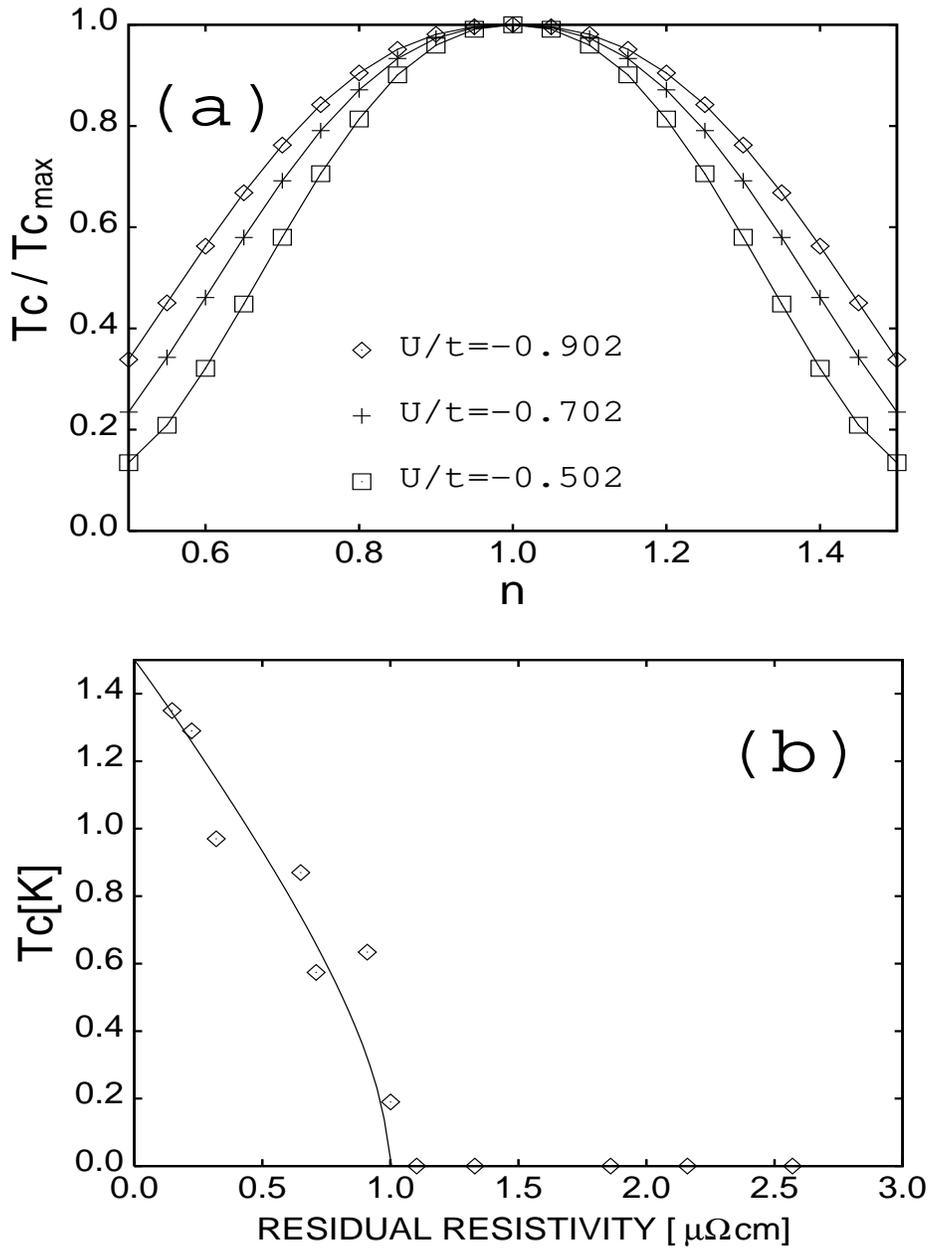

\caption[Fig. 2]{($a$) T$_C$ for the clean system
versus band filling $n$ for various interactions $U$. ($b$) T$_C$ versus residual
resistivity
fitted for Sr$_2$RuO$_4$. The diamonds are the data of Ref. [1]. }
\end{figure}   

\begin{figure}[htb]
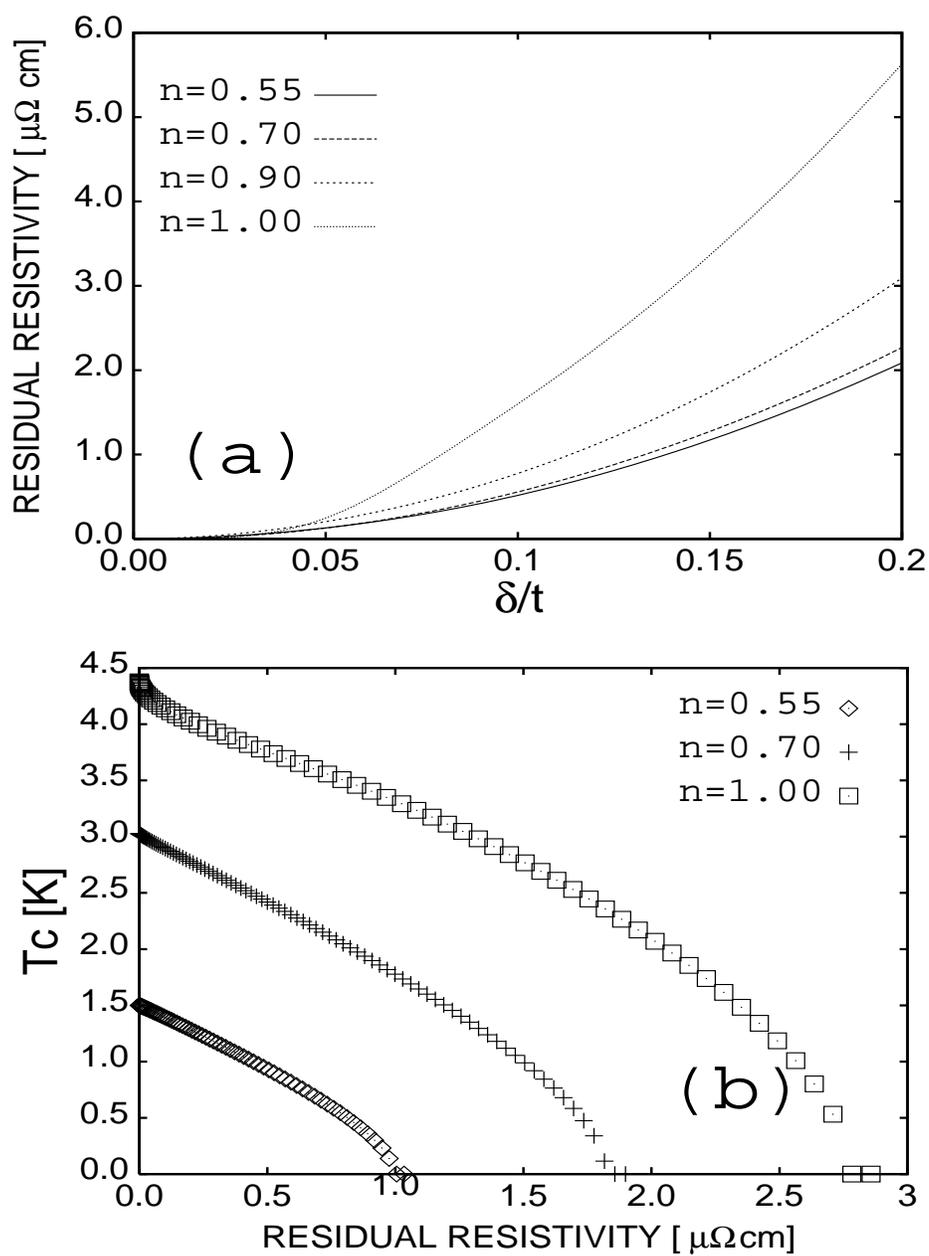

\caption[Fig. 3]{ ($a$) Residual resistivity versus strength of disordered
potential $\delta$ ($\epsilon_i \in
[-\frac{\delta}{2},\frac{\delta}{2}]$)
for various band fillings $n$. ($b$) T$_C$ versus residual resistivity
for various band fillings $n$.
}
\end{figure}

\end{document}